\documentclass[preprint, 3p]{elsarticle}

\usepackage[T1]{fontenc}
\usepackage{hyperref}
\usepackage{lineno}
\modulolinenumbers[1]
\usepackage{amsmath, amssymb, amsfonts}
\usepackage{graphicx, float}
\graphicspath{{img}}

\begin{document}






%

\begin{center}
{\Large  Evolution of polygonal crack patterns in mud when subjected to repeated wetting-drying cycles }\\ \vskip 0.5cm
{ \small
Ruhul A I Haque$^{a,b}$, Atish J. Mitra$^{c}$, Sujata Tarafdar$^b$ and Tapati Dutta$^{a,b,*}$  \\
\vskip 0.5cm
$^{a}$ Physics Department, St. Xavier's College, Kolkata 700016, India\\
$^{b}$ Condensed Matter Physics Research Centre, Physics Department, Jadavpur University, Kolkata 700032, India\\
$^c$ Mathematical Sciences, Montana Tech, 1300 W Park St, Butte, MT 59701, United States \\
* Corresponding Author, Email:tapati$\_$mithu@yahoo.com\\
}
\end{center}
\vskip 1cm

\noindent {\bf Abstract}\\ 
\noindent
 The present paper demonstrates how a natural crack mosaic resembling a  random tessellation evolves with repeated ‘wetting followed by drying’ cycles. The natural system here is a crack network in a drying colloidal material, for example, a layer of mud. A spring network model is used to simulate consecutive wetting and drying cycles in mud layers until the crack mosaic matures. The simulated results compare favourably with reported experimental findings. The evolution of these crack mosaics has been mapped as a trajectory of a 4-vector tuple in a geometry-topology domain. A phenomenological relation between energy and crack geometry as functions of time cycles is proposed based on principles of crack mechanics. We follow the crack pattern evolution to find that the pattern veers towards a Voronoi mosaic in order to minimize the system energy. Some examples of static crack mosaics in nature have also been explored to verify if nature prefers Voronoi patterns. In this context, the authors define new geometric  measures of Voronoi-ness of crack mosaics to quantify  how close a tessellation is to a Voronoi tessellation, or even, to a Centroidal Voronoi tessellation.




\section{Introduction}
Crack patterns (mosaics) are abundant in nature - from microscopical biological structures to huge volcanic columnar joints - they can be found every where\cite{Honda1983, weaire1984, Hayashi2004, Thompson2021, Valentina2022}. The beautiful pattern formed by the outlines of the veins on the wings of a dragonfly or the irregular brown spots on the fur of a giraffe, are all crack mosaics, but of a special nature; these are Voronoi-like mosaics. There is often some kind of optimization involved when a mosaic is Voronoi. For example, the pattern formed by the honeycomb cells, which is actually a Centroidal Voronoi, is made in such a way so as to optimize the use of the construction material - wax \cite{Hales2001}. 

Mathematically, a Voronoi diagram partitions a plane into few regions around a set of points (generators). The regions are known as Voronoi regions and the generator points are the seeds. The Voronoi regions are such that every point in the region is closest to the seed it belongs to. This particular feature makes a Voronoi diagram interesting and it has widespread applications in physical systems. In real systems,  the seeds are the important points  that are expected to control the entire process of tessellation. For example, when the physician John Snow was investigating the September 1854 cholera outbreak in Soho that killed $10\%$ of the population and wiped off entire families in days, he mapped the regions of outbreak to show that the largest area of infection constituted a Voronoi region with the contaminated Broad Street water pump at the seed \cite{Snow1855}. A new approach in
hydrographic geomorphometry involves using Voronoi diagrams generated from airborne laser altimetry data points, to determine flow direction and define watershed boundaries specially in low relief regions \cite{Peters2014}. Another new concept called a `boat-sail distance' has been introduced on the surface of water with flow, to define a generalized Voronoi diagram in such a way that the water surface is partitioned into regions belonging to the nearest harbours with respect to this distance \cite{Tetsushi2003}. 

Dynamical systems like surface cracking due to permafrost,  and columnar joints, evolve over periods of decades and mature into Voronoi-like pattern \cite{Degraff1989, Budkewitsch1994, Aydin1998, Goehring2008, Sletten2003}. In the rough Voronoi tessellation of mud cracks in summer, the seeds of the Voronoi regions are expected to be the local stress points due to desiccation; while in the giant columnar joints formed during the cooling of basaltic lava, the temperature of the individual column is highest at the seeds of the Voronoi regions that are prominent in their transverse cross-section. It has been reported that columnar joints form so as to release maximum thermal stress in order to minimize the system energy \cite{Jagla2002, Ruhul2022}. The abundance of Voronoi tessellation in efficient network systems in different spheres of life, prompts a closer inspection of the fundamental driving forces responsible for such a geometry.  

In this paper the authors report the evolution of random mud cracks under repeated cycles of wetting followed by drying. Mud slurries are colloidal systems that are made up of large particles with a long relaxation times, often $\approx 10^{10}$ times higher than other molecules. In fact - clay, mud, lava - are all examples of systems whose dynamical maturation  could be tracked by scientists only because their slurries have long relaxation times, typical of colloidal systems. This has prompted the authors to simulate the time evolution of crack patterns in wetting-drying cycles of desiccating clay systems using a spring network model.

 The initial geometry of desiccation crack mosaic changes with increasing cycles till it reaches a state of maturation. In an earlier work the authors have proposed a topology-geometry combinatoric $(n,v,D,\lambda)$ that is characteristic to a  crack mosaic \cite{Anamika2022}; here $(n, v)$ describe the topology and $(D, \lambda)$, the geometry of the crack mosaic. Greater details of these topological and geometrical measures are described in the following section. The description of any crack mosaic by its $(n,v,D, \lambda)$ quadruple enables the representation of the mosaic as a unique point in a 4-dimensional topology-geometry domain that describes a subset $\mathbb{R}^4$. Thus crack mosaics having similar $(n,v,D,\lambda)$ points show up as point clusters in the domain. Real crack meshes are mostly clustered around the Gilbert
crack representative point with only a few representative points of crack mosaics clustered around the Voronoi crack descriptor. Much of the mathematically allowed $(n,v,D, \lambda)$ domain space remains empty. As wetting-drying cycles in mud cracks are reported to start with an abundance of T-junctions (Gilbert mosaics), and mature towards Voronoi-like mosaics with Y-junctions, the authors have tracked the evolution of the topology and geometry of these crack mosaics  as a trajectory in the $(n,v,D,\lambda)$ domain. Thus the development of a complex dynamical system over time, is effectively comprehended and may be characteristic of such systems. 

In an effort to study and quantify the `Voronoi-ness' of a physical tessellation, the authors introduce a $p$ measure which can provide a heuristic quantification of how `Voronoi-like' is the tessellation. Further, the authors have suitably utilized the Hausdorff Metric $d_H$, to measure the deviation of any crack mosaic from a Centroidal Voronoi. 
 Quantification of the `Voronoi-ness' in terms of $p$ and the Hausdorff Metric, as a function of time cycles is done on the shape evolution of the crack mosaic. The patterns obtained from our simulations are very similar to real system observations reported by physicists and geologists. The authors  further define the geometry of the mosaic in terms of the geometric measure `iso-perimetric' ratio $\lambda$, and propose a phenomenological relation between $\lambda$ and change of fracture energy, where both $\lambda$ and the energy are function of wetting-drying cycles.  It is observed that the time evolution of desiccation mud crack mosaics tend towards a Voronoi tessellation as this minimizes the total energy of the system.

The distribution of seeds of a Voronoi region controls the shape and size of the associated regions of the tessellation. Hence identifying the seed of a Voronoi region can determine the range of correlation of a particular physical property, e.g. elastic properties of surface cracks. In the case of Centroidal Voronoi tessellations, the seed and the centroid of the corresponding Voronoi region coincide. However  in real crack mosaics the seeds cannot always be determined with certainty, and the centroid of the polygon is often approximated to act as the seed. Hence it remains a matter of interest to identify not only the extent of `Voronoi-ness' of any polygonal network, but also, how close is the Voronoi to a Centroidal Voronoi.
The authors show that the $p$ measure and the Hausdorff Metric to the corresponding Centroidal Voronoi mosaic, together, can determine (a) how Voronoi-like is a crack mosaic, and (b) how close is a Voronoi region to being a Centroidal Voronoi. 

In the following sections, we first provide a simple explanation of the mechanics of desiccation cracks.  The simulation model of wetting and drying dynamics of a clay system is described next in terms of a spring network model based on the mechanics of desiccation cracks. The 4-tuple combinatoric describing the topology and geometry of crack patterns is introduced and explained next.  This is followed by the introduction of the phenomenological relation between fracture energy and crack geometry based on heuristic arguments based on the fundamental principles of fracture mechanics. The energy changes that occur with increasing drying cycles is further examined from the aspect of the mathematical definition of geometrical energy of a mosaic. Measures for the quantification of `Voronoi-ness' of a crack mosaic are introduced next. Having established the fundamental theories guiding this work, results of the simulation are examined in terms of the time evolution of both - the crack mosaic geometry and the system energy. `Voronoi-ness' of the crack mosaic is represented in terms of wetting-drying cycles and quantified in terms of defined measures. The preferred `Voronoi-ness' of desiccating crack mosaics is explained next in tandem to energy minimization of the system. 
An examination of few natural but static crack mosaics is also conducted to check if Nature does indeed have any preference for Voronoi patterns in mosaics.
 The paper concludes with a brief summary of the important aspects of this work.   

\section{Theory}

\subsection{Crack evolution}
Most crack patterns observed in nature are static, i.e., their shapes do not change over time. Examples of static crack mosaics can range from the almost regular hexagonal pattern of honeycombs to the disordered polygonal mud cracks of dry agricultural fields and river beds. However if the time scale of changes in crack pattern is of the order of experimental/observation time, it is possible to track the dynamic evolution of crack patterns upto maturation.  There are few examples of natural crack mosaics that show dynamic changes in shape to eventually mature to Voronoi-like pattern - for example, terrain cracks due to permafrost, columnar joints, patterns on salt deserts or dry salt lakes \cite{Lasser2023} etc.
Similar nature of crack evolution is observed when wet corn starch is dried \cite{Muller1998a, Muller1998b, Goehring2009}. Laboratory based experiments on desiccating mud cracks subjected to repeated wetting-drying cycles have been reported where the angles between cracks change from $90^\circ$ to $120^\circ$ \cite{Goehring2010, Goehring2013} -  which leads us to wonder whether the time evolution of such crack patterns tend towards Voronoi-like tessellations.  To develop a better understanding of the fundamental reasons driving such evolving crack systems, we simulate a repeated wetting-drying experiment on mud-crack using spring network \cite{Sadhukhan2011}. We examine the `Voronoi-ness' of the dynamic crack pattern in terms of the measures described in the following section.

During desiccation of mud or other colloidal systems, the capillary pressure in between the particles increases with evaporation and the particles are drawn closer together \cite{Pauchard2018}. Micro-cracks initiate from inhomogeneities that act as crack seeds \cite{ST2015_book}. If a new crack appears in the vicinity of an existing crack, it grows to join the existing crack at right angles.  This is explained by the fact that stress is higher parallel to the existing crack and it is therefore energetically preferable that the growing crack bends along the direction of highest gradient of stress release, i.e. meets the existing crack normally \cite{Bohn2005}. As a result a highly connected crack network is formed dividing the entire mud plane into rectilinear regions (peds). As crack faces open up, they too contribute to surface evaporation. Thus in every crack ped a moisture gradient is created that increases from an exposed crack face towards the centre of the ped, i.e., the drying front of each ped in the mosaic proceeds from the crack boundary towards the centre. As drying continues, the cracks broaden as the ped area shrinks. If the cracked surface is then wetted till saturation, the cracks spaces and the inter-particle pores between colloidal aggregates fill up with water, i.e. the elastic strain generated during desiccation is relaxed a little. In the case where the mud/clay system is wetted and dried repeatedly, the vertices where two different cracks intersect, begin to shift slightly and the angles produced at the joints gradually change from $90^\circ$ to $120^\circ$. The `T' like junctions transform to `Y' like junctions and the cracks pattern tends towards hexagonal.

\subsection{Combinatorial topology and geometry of crack mosaics}
The topology-geometry combinatoric $(n,v,D,\lambda)$ is characteristic to a  crack mosaic \cite{Anamika2022}, where if the node of the crack mosaic is the vertex of $n$ polygonal cells or crack `peds', the node is said to have degree $n$.  A cell having $v$ vertices or edges, is assigned a degree $v$. Most convex polygonal crack meshes  in nature show $(n,v)$ values ranging from $(2,4)$ to $(4,4)$, depending on whether majority of the nodes are irregular or regular respectively \cite{Domokos2020}. The $(2,4)$ topological description of a crack mesh is the signature of the Gilbert tessellation \cite{Gray1976, Mackisack1996}. Voronoi cracks show a topological combinatoric of $(3,6)$ \cite{Dirichlet1850, Voronoi1908}.
In general $(2,4)$ and $(4,4)$ indicate rectangular and $(3,6)$ indicates hexagonal geometry of the lattice.

The measure of the geometric regularity of the entire mosaic is given by $D$ defined by $
 \displaystyle{D=\frac{\sum_{N=1}^M D_N}{M}}$ 
  where $M$ is the total number of polygons (cells) in the mosaic; $D_{N}$ is the angular defect of the $N^{th}$ polygon of the mesh defined by
  
  \begin{equation}
 \displaystyle{D_N=\frac{1}{1 + \sum_{i=1}^{\textsf{v}_N} \lvert \theta_i -\frac{(\textsf{v}_N-2)\pi}{\textsf{v}_N}\rvert}}
 \label{DN}
\end{equation}
 where $\textsf{v}_N$ is the number of vertices of the $N^{th}$ polygon. 
 
 In words, $D$ is the angular deviation, implying the average difference between the angles of the polygon and the angles it would have for a regular polygon with the same number of sides. By construction, the measures $D_N$ and $D$ range in $[0,1]$. Another useful geometric measure is $\lambda$, the normalized isoperimetric ratio defined by $ \displaystyle{\lambda = \frac{\sum_{N=1}^M \lambda_N}{M}}$, averaged over all polygons in the mosaic. Here $\lambda_N$ is isoperimetric ratio of the $N^{th}$ polygon  defined by
 
 \begin{equation}
 \displaystyle{\lambda_N = \frac{4 \pi A_N}{l_N^2}}
 \label{lambda}
\end{equation}  
Where $A_N$ and $l_N$ are respectively, the  area  and perimeter of the $N^{th}$ crack ped of the mosaic. 
 Thus the isoperimetric ratio is a dimensionless quantity that quantifies the shape of a polygonal ped in the crack network. So $\lambda$ varies from 1 towards  0 as the structure of the polygons of the network  changes from relatively circular towards more and more elongated shapes. 
 
 Inspecting the mathematically allowed space of $(n,v,D,\lambda)$, it is observed that real crack meshes are mostly clustered around the Gilbert crack representative point with only a few representative points of crack mosaics clustered around the Voronoi crack descriptor;  much of the allowed domain space remains empty of real crack mosaic points. In this paper we demonstrate how randomly oriented crack mosaic showing Gilbert tessellation evolve with wetting-drying cycles towards a Voronoi tessellation at maturation.
 
\subsection{Fracture Energy}
Though the crack evolution happens in the horizontal plane of the top surface of the mud layers, wetting-drying cycles have their effects up to a few layers beneath the top surface. Thus the system can be thought of as a quasi 2-D system with thickness $z$, within which the major energy changes occur. It is assumed that the surface crack penetrates the thickness $z$ in infinitesimal steps  of $\delta z$ in a single time cycle, henceforth referred to as cycle step $t$. The total energy of the system at any cycle step can be divided into two contributing parts -- (a) elastic energy of the peds (b) the fracture energy spent in creating new crack area. The elastic energy of the $i^{th}$ ped depends on its area $A_i$ and thickness $\delta z$ and is given by $ \gamma\sum_{i}A_i\delta z$; the parameter $\gamma$ represents the characteristics of the mud system. The fracture energy to create a fresh surface of  $L\delta z$ in a single cycle step is $ \eta L\delta z$, where $\eta$ is the fracture energy per unit area and $L$ is the total crack length. Then the total energy of the system per unit height is 

$$ E = \gamma\sum_{i}A_i  + \eta L $$
The cracks being very thin compared to the ped area, total area of the surface is assumed to remain constant throughout the simulation. Only the shapes of the polygons change. Therefore it is convenient to express the energy in terms of the shape parameter, i.e., the iso-perimetric ratio of the tessellation, $\lambda$. The energy can be expressed in terms of $\lambda$ as follows \cite{Ruhul2022} --
\begin{equation}
E = \alpha \lambda^{-\beta}
\label{energy}
\end{equation}
where $\alpha, \beta$ are constants that capture  ambient effects and the physical characteristics of the system respectively. 

\subsection{Geometrical Energy} A function $\varepsilon_i$ can be defined for each of the polygonal regions as $\varepsilon_i =  \int_{x \in \Omega_i} \rho(x) \|x_i - x\|^2 dx$, where  $\|\cdot\|$ is the Euclidean norm in $\mathbb{R}^2$, $\rho(x)$ is a density function, $\{x_i\}_{i=1}^n$ are points on the polygonal plane and the integration is over the entire area of the polygon $\Omega_i$. If $\varepsilon_i$ is interpreted as the strain energy of the $i^{th}$ polygon, then the stress density function $\rho(x)$ is obtained from a linear stress distribution inside the polygon such that the stress is maximum at the $\rho$-weighted centroid of the polygon (which is almost identical to the geometric centroid with a deviation $< 0.01\%$) and zero at the boundaries.

A polygon in the mosaic shrinks as a result of the strain developed due to the moisture gradient in a ped  as drying commences from the edge towards the centre. This implies that stress due to shrinkage is zero at the crack front and highest at the centroid. Hence a linear stress distribution in a crack ped is a simple approximation that mimics the stress distribution in the ped. It is assumed that stress due to evaporation from the exposed top layer is a constant for the whole mosaic. The geometrical energy function of the tessellation is then defined as the summation of this function for all polygons in the tessellation:
  $$\varepsilon = \sum_i \varepsilon_i = \sum_i \int_{x \in \Omega_i} \rho(x) \|x_i - x\|^2 dx $$
A necessary condition for this energy function to be minimum is that the tessellation must be a Centroidal Voronoi tessellation with $\{x_i\}_{i=1}^n$ as the seeds.

\subsection{Quantification of Voronoi-ness} \label{sec-quntification}
Given a finite number of points $x_1, x_2 \dots x_n$ on a subset $\Omega$ of the plane, we define $\Omega_{i} = \{ x \in \Omega : \|x - x_i\| \leqslant \|x - x_j\| \text{ for all } j \neq i\}$. A Voronoi diagram with seeds $x_1, x_2 \dots x_n$ is the mosaic of $\Omega$ by the regions $\Omega_1, \Omega_2 \dots \Omega_n $. This provides a polygonal mosaic of $\Omega$ with interesting geometric properties \footnote{A related construction, the Delaunay Triangulation, is a triangulation of a finite set of points $F$ with the Empty Circumcircle Property (no point in $F$ is contained in the interior of any triangle of the triangulation).}. A Voronoi diagram is the dual construction of the Delaunay triangulation of the same set of points. For an introduction to the mathematical properties of Voronoi diagrams and Delaunay triangulations in this section and elsewhere in this work, see  \cite{Okabe2009}.
Appendix provides a step-by-step diagrammatic flow chart for the construction of a Voronoi diagram from a scatter of seeds.

Given a mosaic of the plane it is in general a difficult problem to ascertain if it is a Voronoi diagram for a set of points, the difficulty being that no set of seeds is pre-determined. There are mathematical techniques of checking if a mosaic is Voronoi, but those involve a linear programming problem. For real life mosaics, a computationally simpler option is to consider the centroids of the regions of the mosaic as seeds of a Voronoi diagram which can be then compared to the original mosaic.  As an example, in crack mosaics the seed of stress cannot always be determined with certainty.  As an approximation, the centroids of the polygons may be assumed to act as the seeds.  We can then triangulate the mosaic by joining the centroids of the polygons that share a common vertex, Fig. \ref{triangulation}. A defining feature of Delaunay triangulation is that not a single circumcircle of the Delaunay triangles includes any of the seeds. This feature can be used to quantify the fraction of non-empty triangles that do not satisfy the characteristic of Delaunay triangulation. We can define a parameter $p$ (as a heuristic measure of the `Voronoi-ness' of real crack systems) as:
$$p = \frac{\text{Number of non-empty circumcircles}}{\text{Total number of the triangles}} $$

A more definitive way to measure the `Voronoi-ness' is the mathematical tool Hausdorff Metric\footnote{For compact subsets $A,B \subset \mathbb{R}^2$, the Hausdorff Metric $d_H(A,B)$ is defined as $\displaystyle{d_H(A,B)= \max\{\max_{a \in A}d(a,B),\max_{b \in B}d(b,A)\}}$, where $d(x,C)= \inf \{\| x-c \| : c \in C \}$.}. It measures the distance between two subsets of a metric space. We can compare a real crack with the corresponding Voronoi diagram generated by using the centroids of the crack peds as the seeds \cite{Ruhul2022}.

\section{Materials and Method}

\subsection{Simulation of the wetting-drying process}

As  desiccation cracks appear at random and grow along a straight line as mentioned above, we use the Gilbert tessellation \cite{Gilbert1967, Anamika2022}, which is a random network, to model the initial crack pattern. Minute crack tips are initialized from  a homogeneous Poisson point distribution of seeds, and allowed to grow in randomly chosen directions on a $100\times100$ square plane. The cracks begin to grow in both directions. Whenever a growing edge collides with an existing crack segment, it stops growing and forms a node there (\textit{movie of Gilbert crack simulation available as supplementary file}). Once the entire crack network is formed, the nodes and the distinct crack regions, i.e, the peds, are identified.

\begin{figure}[h!]
    \centering
    \includegraphics[width=\linewidth]{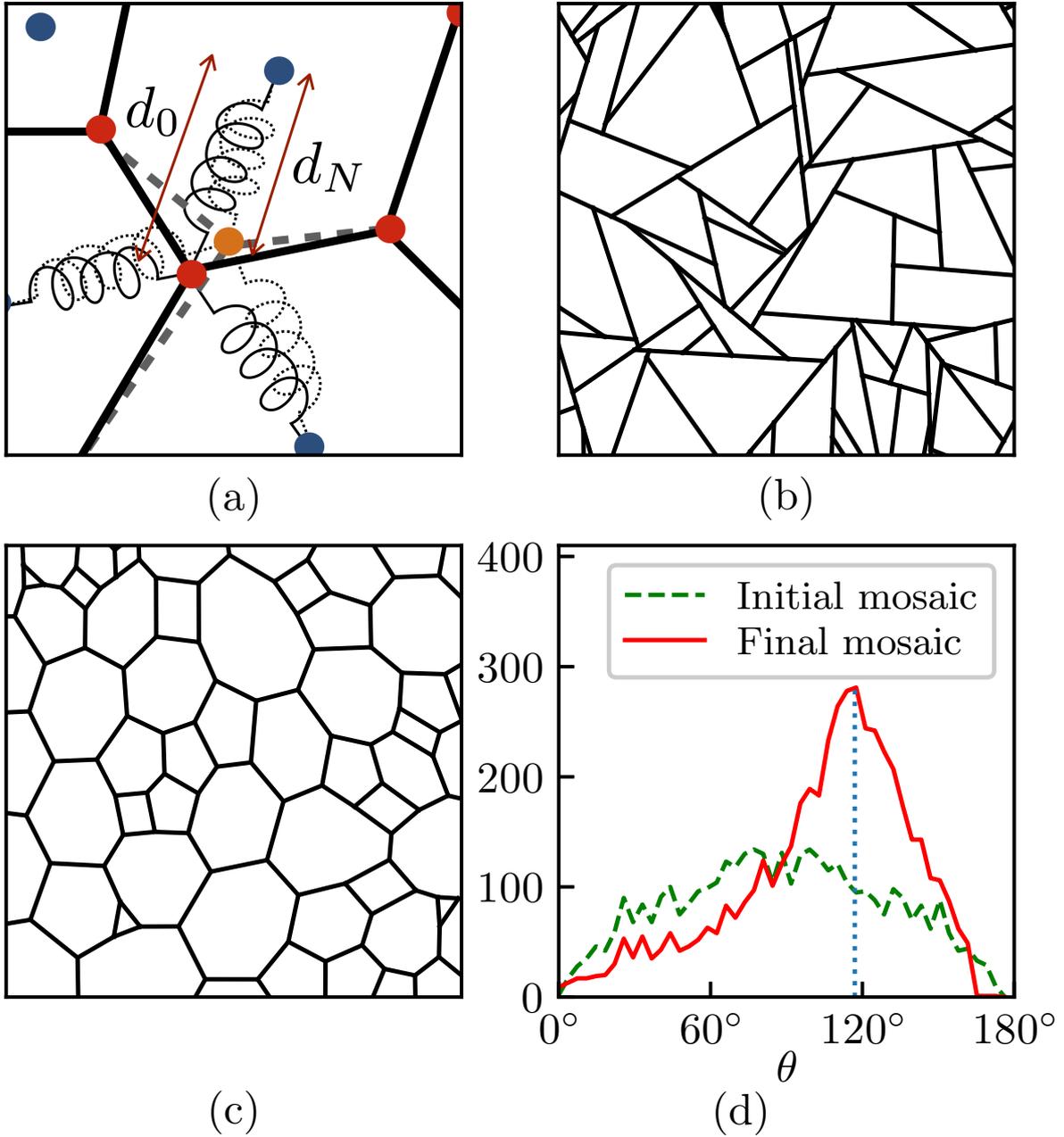}
    \caption{(a) Spring model to simulate desiccation cracking. The crack ped vertices are marked by red circles. The ped centroids are marked by blue circles. $d_{0}$, the natural spring length between centroid (blue) and vertex (red) indicated by arrow.  A red vertex is acted upon by the elastic forces, represented by springs, from its surrounding polygons. The vertex is displaced to a new position (orange circle) after $N$ desiccation steps, where it remains connected to the polygon centroids with springs (shown by dotted springs) of new lengths $d_{N}$ (indicated by arrow). (b) Initial crack network of a desiccation crack, (c) Final matured pattern of the crack network after applying repeated wetting-drying cycles. (d) Initial and final distributions of the polygonal angles of crack peds. The final distribution shows a peak at  $ \approx 120^\circ$, indicated by blue dotted line. }
    \label{schematic}
\end{figure}

 The resulting crack mosaic and the dynamics of repeated wetting and drying of the system, is mimicked by a connected spring network. As explained earlier, the crack faces along the boundary of every ped contributes to surface evaporation. As the drying front of each ped proceeds inwards with desiccation time, we have replicated the process by connecting each  vertex of the polygon (a ped) in the mosaic to its centroid via a spring of spring constant $k$, that gets compressed with drying time steps. The natural length $d_0$ of the spring is taken to be equal to the distance between the vertex and the centroid, Fig. \ref{schematic}(a).\\

 The evaporation process is governed by not only ambient conditions, but all the material properties of the colloidal system. In our simulation,  the natural lengths of the springs shrink due to desiccation according to the following relation --
\begin{equation}
d_n = d_{n-1} \left(1 - \frac{a}{b^n} \right)
\label{drying}  
\end{equation}
where $a$ and $b$ are  parameters that can control the desiccation process and $n$ is the number of time steps of desiccation between subsequent wetting. The desiccation rule was based on earlier works by the authors  Tarafdar and Dutta \cite{Sadhukhan2007, Sadhukhan2011}, where evaporation rate measurements on aqueous clay showed that evaporation usually  continued till the mass got reduced by $\approx 63\%$ of its original weight.  As the system desiccates and the springs shrink, a net force that is the resultant of forces due to all springs connected to that node, is exerted on it , Fig. \ref{schematic}(a). The displacement of any node can thus be expressed in terms of the shrinkage of the spring length as follows --
\begin{equation}
\Delta \vec{x}  =  \kappa\sum_i \Delta \vec{d_i}
\label{node-displacement}
\end{equation}
where the summation runs over all the springs that are connected to the node.  $\Delta \vec{d_i}$ is the shrinkage of the $i^{th}$ spring in a single cycle and is given by $( d_{0} -  d_{N})\hat{n}$.  Here $d_{N}$ is the length of the spring after desiccating for $N$ time steps, before the next bout of wetting is commenced.  $\hat{n}$ is the unit vector directed towards the centroid.  The proportionality constant $\kappa$ is a function of the physical parameters $a$ and $b$ of the system. The crack evolution during the drying cycle is mimicked by displacing the nodes due to the resultant forces, and thereby changing the paths of the crack segments. 

During the wetting process the springs expand  and their natural lengths change with time as follows  --
\begin{equation}
 d_{n^\prime} = d_{n^\prime-1} \left(1 + \frac{a}{b^{n^\prime}} \right) 
 \label{wetting}
\end{equation}   
The nodes move in a similar fashion as in the the drying process. However as all wetting-drying experiments are performed with wetting the sample till saturation, we have kept the number of wetting steps $N^\prime=1$ to indicate saturation during wetting.  At the start of wetting-drying cycle, i.e., at $t=0$, $n=1$ and $d_{n-1} = d_0$ in Eq.(\ref{drying}). After $N$ number of drying steps all springs connected to a single node shrink according to Eq.(\ref{drying}) and the node is displaced following Eq.(\ref{node-displacement}). All nodes are updated to their new positions in parallel. The $d_0$ value of every spring is reassigned  with respect to its new position to its centroid. Wetting and its following spring expansion is commenced according to Eq.(\ref{wetting}).  A single wetting-drying time cycle $t$ is made up of $N$ successive drying steps followed by $N^\prime =1$ wetting step. (We note here that the recursive relation in Eq. \ref{drying}, with $b > 1$ and $a > 0$ forms an infinite product which converges to a positive real number. This convergence corresponds physically to sufficiently many drying cycles.)  At the end of every time cycle $t$, the centroids of the mosaic polygons are determined. The centroids act as seeds for the construction of a Voronoi mosaic. The Hausdorff Metric between the the Voronoi mosaic and the crack mosaic is computed for every cycle $t$. 
 The cycle of these wetting and drying processes is repeated until the difference in the displacement of any vertex at consecutive time steps is $\leq 10^{-3}$. The crack pattern is then assumed to have matured and stops evolving further. (\textit{movie of crack node movement due to wetting drying cycles is available as supplementary file})

\section{Results and Discussion}
In the following subsections, we first present the results and discussion on  dynamical evolution of mud cracks under repeated wetting-drying cycles from our simulation. The topological and geometrical changes of the crack mosaic are mapped onto a trajectory in the $(n, v, D, \lambda)$ domain. The preference for `Voronoi-ness' at maturation is explained from the computation of energy evolution of the system towards a minimum. The $p$ and Hausdorff metric evolution with time cycle $t$ are examined and discussed. Lastly we present our analysis on some natural crack mosaics, i.e., matured crack patterns, to demonstrate that energy minimization principle is one of the major driving forces behind preference for Voronoi geometry in crack mosaics. 

\subsection{Dynamic progression of crack mosaics}
Figure. \ref{schematic}(b) shows the initial desiccating crack network and Fig. \ref{schematic}(c), its matured state after repeated wetting-drying cycles. The figure has been generated with parameter values of $a = 0.05$, $b = 1.2$ and $N = 25$. The simulation has been run to check the effect of ambient by varying $a$, the effect of the physical properties of the system by varying $b$, and finally by changing the periodicity of the wetting-drying cycles by varying $N$. An increase in $N$ value signifies longer intervals between successive wetting events. 
\begin{figure}[h!]
\centering
\includegraphics[width=\linewidth]{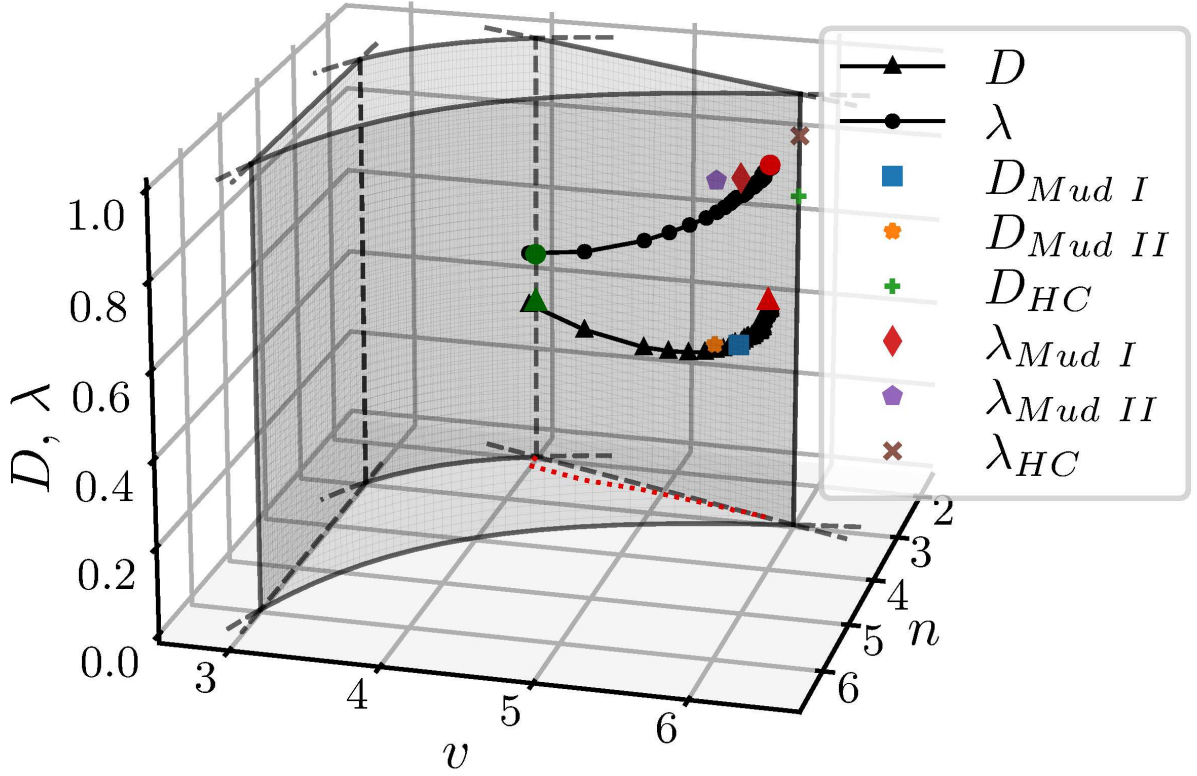}
\caption{Trajectory of the crack evolution under repeated cycles of wetting and drying in the $(n, v, D, \lambda)$ domain. The position of the static crack mosaics of (i)Mud crack I (ii) Mud crack II and (iii) Honeycomb, in the 4-tuple domain are also indicated as explained in the legend. }
\label{pattern-gil-uni-3d}
\end{figure}
Figure. (\ref{pattern-gil-uni-3d}) displays the time development of the topology and geometry measures for the simulation. The $(n,v)$ values increase sharply from $(2,4)$ to $(3,6)$ within the first few time-steps, and thereafter remain almost constant in time. The geometric measure $D$  decreases sharply at first, and then increases to a constant following a power-law function. The trajectory is a map of the crack mosaic under repeated wetting-drying cycles from a Gilbert tessellation of $(n,v) = (2,4)$ towards a Voronoi tessellation of $(n,v) = (3,6)$ upon maturation.

\begin{figure*}
    \centering
    \includegraphics[width=0.95\textwidth]{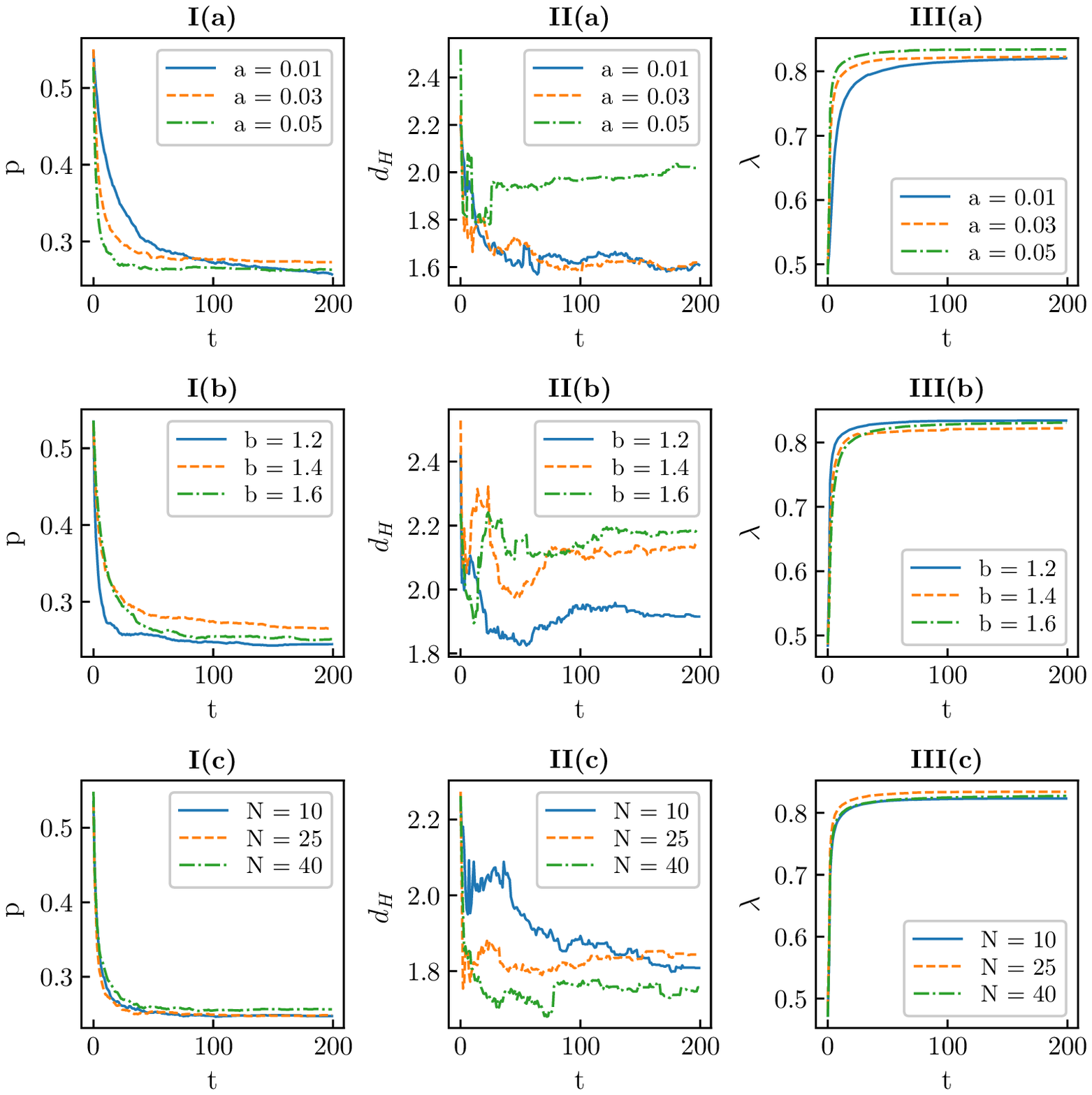}
    \caption{Column I(a--c): The ratio of the  circles that include any seed, $p$ decreases with wetting-drying cycles. Column II(a--c): Change in Hausdorff Metric of the mud-crack and the corresponding Voronoi diagram generated using the centroids. The metric decreases with the wetting-drying cycles and saturates at a constant value around $1.8$.  Column III(a--c): The Iso-perimetric ratio $\lambda$ increases as the mud undergoes wetting-drying cycles. This indicates that the peds are becoming rounder with wetting-drying cycles. In the first row (I(a)-III(a)) the parameter `$a$' is only varied while keeping the other parameters fixed. In the second and third rows the parameter `$b$' and the number of desiccating time steps $N$ in a single cycle are varied respectively while keeping the other parameters fixed. } 
    \label{hausdorff_t}
\end{figure*}

We have investigated the role of the parameters $a$, $b$ and $N$  that represent respectively - the ambient effect, the rate of spring shrinkage that essentially encapsulates material properties, and rate of cycles, on the evolution of the crack mosaic. The final rate at which any mosaic under repeated wetting drying cycles will mature towards a Voronoi diagram, is an optimization of all the three parameters. Fig. \ref{hausdorff_t} demonstrates the effect of these parameters on $p$ and Hausdorff Metric  measures of the mosaic. The fact that both $p$ measure and the Hausdorff Metric distance (of the crack mosaic with the corresponding Centroidal Voronoi mosaic) decrease with $t$ is an indication that the crack mosaic veers towards a Centroidal Voronoic pattern with increasing number of cycles. In the matured mosaic all nodes are trivalent. A large proportion of nodes form a $120^\circ$ triad, Fig. \ref{schematic}(d). Our algorithm geometrically ensures that in such cases we will get the Voronoi empty circle condition, which agrees with our simulations.
Fig. \ref{hausdorff_t} column I(a--c), shows that $p$ is lower for higher $a$ and lower $b$ values, but the effect of $N$ is not very distinct. A longer drying spell before wetting (higher $N$) and a lower elastic modulus (lower $b$), as indicated by Fig. \ref{hausdorff_t}(b--c), is conducive towards quicker approach to Voronoi nature. 

 For a regular hexagon the value of $\lambda$ is 0.907. From our simulation results it can be seen that the value of $\lambda$ increases  from around 0.45 and saturates to 0.83 reflecting the approach to a nearly hexagonal shape, Fig. \ref{hausdorff_t} column III(a--c). This is corroborated from Fig. \ref{schematic}(c) which displays the matured crack mosaic. Every polygon appears `rounder' than its initial geometry in the matured mosaic.
In brief, our simulations indicate that our algorithm - irrespective of the initial crack mosaic -  seems to converge to a Centroidal Voronoi mosaic. Moreover, in our simulations a material with lower elastic modulus when subjected to longer drying spells between consecutive wetting, is conducive for the system to reach a mature Centroidal Voronoi-like mosaic quickly. From the time evolution of the measures of both $p$ and the Hausdorff Metric, it is evident that periodic wetting and drying drives desiccation crack mosaics towards more Centroidal Voronoi-like.

\begin{figure}[h!]
    \centering
    \includegraphics[width=\linewidth]{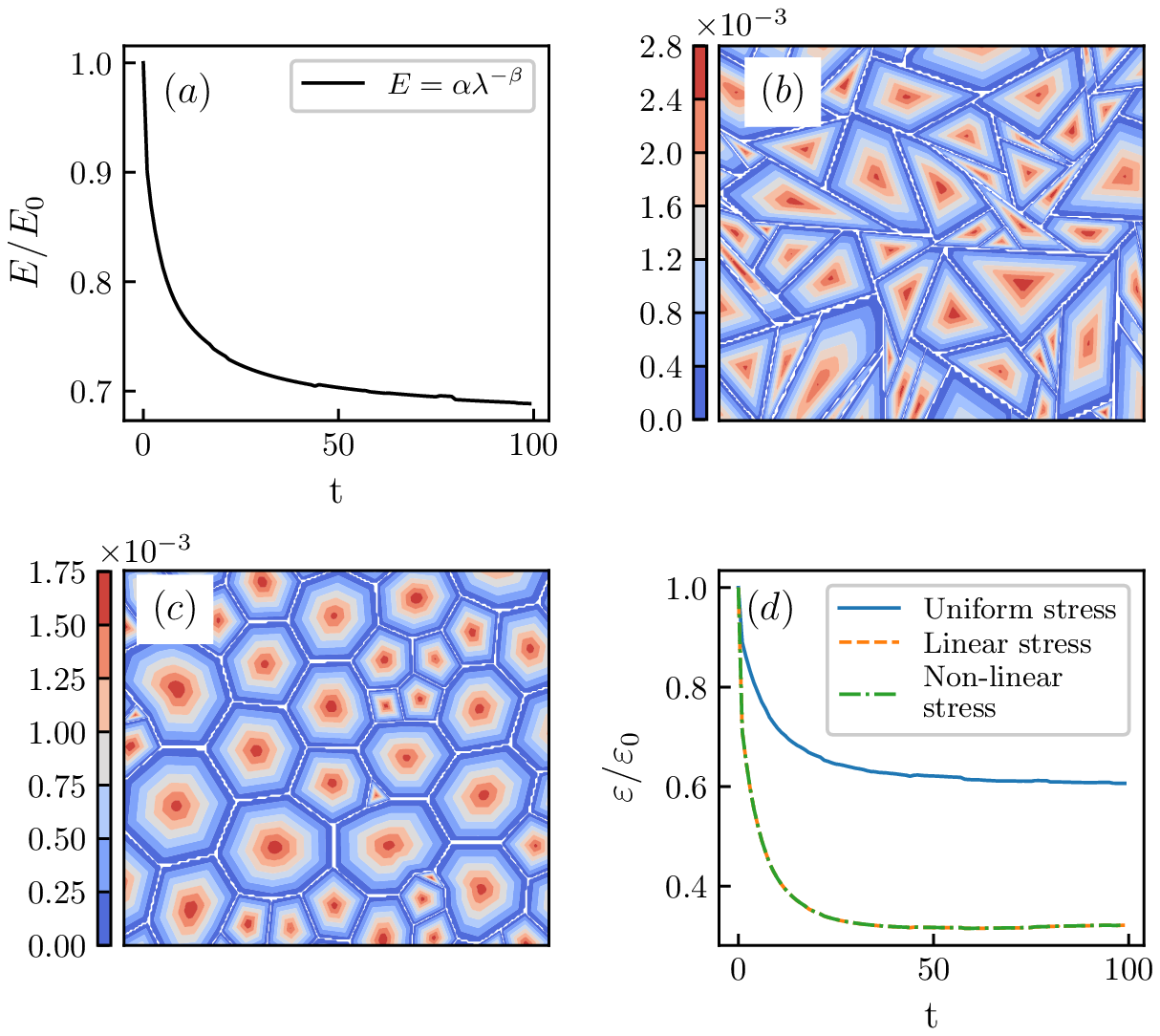}
    \caption{Energy minimization of (a) Elastic energy with wetting-drying cycles. Linear stress distribution inside each of the polygons of the (b) initial and (c) final mosaics. (d) Geometrical Energy for different stress distributions inside a polygon}
    \label{energy_vs_t_norm}
\end{figure}

The geometrical parameter $\lambda$ and total crack length $L$ are computed at every cycle step $t$, Eq.(\ref{energy}),  and the the net system energy change in terms of the crack geometry parameter $\lambda$ is plotted in Fig. \ref{energy_vs_t_norm}(a). Thus the system energy has a power-law dependence on $\lambda$ with the exponent $\beta = 1.13$. It is apparent that the energy minimizes with time and saturates to a constant value as the crack pattern matures and becomes Voronoi.
 Figs. \ref{energy_vs_t_norm}(b,c) display the stress distribution in every ped of the mosaic at the commencement of wetting -drying cycle upto the crack mosaic maturation. The stress distribution shows an approximately linear distribution about the ped centroid following the crack boundary geometry.
Fig. \ref{energy_vs_t_norm}(d) shows the change in the geometrical energy as the crack mosaic evolves with the wetting-drying cycles. It seems that the energy tends to a minimum value indicating that the time-evolution of the tessellation is tending towards a Centroidal Voronoi tessellation.
\par If the polygons in the mosaic have some inhomogeneities by way of micro-cracks or defects,  $\rho(x)$ can be a non-linear function of $x$. Fig.(\ref{energy_vs_t_norm}d) displays that despite the presence of non-linearity in $\rho (x)$, the geometrical energy of the mosaic converges towards an identical equilibrium minimum. With a constant $\rho(x)$, the system energy decreases  following similar trend as before but shows a higher equilibrium minimum value. This suggests that the Voronoi diagram is the equilibrium geometry preferred by dynamically evolving system.

\subsection{Static crack mosaics}

\begin{figure}[h!]
    \centering
    \includegraphics[width=\linewidth]{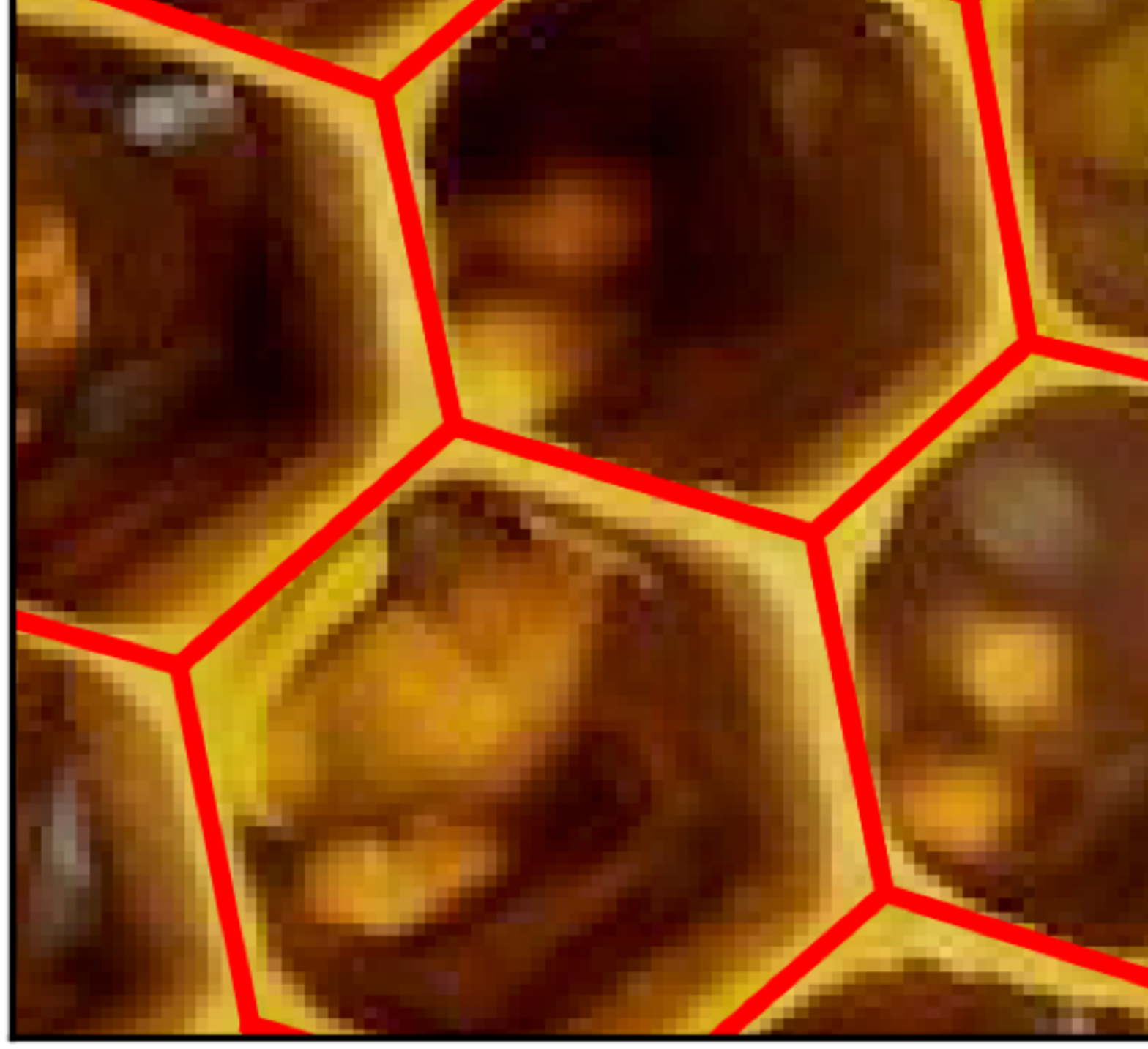}
    \caption{Checking for Voronoi-like patterns. Real mud-cracks: (a) Case I, the red lines outline the crack network. (b) Triangulated by joining the centroids of the polygons that share a common edge. (c) Circumcircle for each of the triangles is drawn. The red circles contain one or more centroids inside them. (d) Case II of mud cracks. (e) and (f) show Delaunay triangulation and circumcircles for all the triangles respectively. Not a single centroid is inside the circles in this case. (g)--(i) show the same schematic flowchart for honeycomb structure. }
    \label{triangulation}
\end{figure} 

To understand if nature does indeed prefer Voronoi patterns in crack mosaics, we study a few examples of real crack mosaics that are static. Fig. \ref{triangulation}(a) and (d), show pictures of actual mud crack mosaics, while Fig.  \ref{triangulation}(g) shows the pattern on the surface of a honeycomb. The red lines outline the polygonalized crack networks. The crack plane is triangulated by connecting the centroids, marked by blue circles, of the polygons  that share a common vertex, Fig. \ref{triangulation}(b, e and h). Circumcircles to each triangle have been constructed as shown in Fig. \ref{triangulation}(c, f and i). We have marked in green those circumcircles that satisfy the condition of Delaunay triangulation. If a triangle is not a Delaunay triangle, the corresponding circle is marked red. Comparing Fig. \ref{triangulation}(c) and Fig. \ref{triangulation}(f) it is evident that though both arise from mosaics of mud cracks, the former triangulation has some non-Delaunay triangles (the red circles are indicative) while the latter  triangulation forms a Delaunay triangulation (all green circles). When the similar exercise is done on a the honeycomb structure of Fig. \ref{triangulation}(g), a Delaunay triangulation is obtained.

\begin{table*}
    \caption{Measures of Voronoi-ness and $(n, v, D, \lambda)$ values for the static crack mosaics.}
    \begin{center}
        \begin{tabular}{c c c c c c c} \hline
             \textbf{Model} & $p$ & $d_H$ (pixels) & n & v & D & $\lambda$ \\ \hline
             Mud crack I & 0.077 & 21 & 2.87 & 5.42 & 0.387 & 0.774 \\ 
             Mud crack II & 0.00 & 14 & 2.95 & 5.60& 0.399 & 0.791 \\ 
             Honeycomb & 0.00 & 0 & 3.00 & 6.00 & 0.765 & 0.904 \\ \hline
         \end{tabular}
    \end{center}
    \label{tab_quantification}
\end{table*}

The $p$ values for all three crack mosaics are displayed in Table \ref{tab_quantification}. The non-zero $p$ value of Fig. \ref{triangulation}(a) establishes that the mud crack  does not form a Centroidal Voronoi diagram. The fact that we get $p = 0$  for Figs. \ref{triangulation}(d and g) tells us that we have a Delaunay triangulation whose dual Voronoi diagram however, may not coincide with the original crack. Thus $p$ is a simple heuristically defined measure which can be computed easily and provides a quick but only partial information about Centroidal `Voronoi-ness'.
 \\
\begin{figure}[h!]
    \centering
    \includegraphics[width=\linewidth]{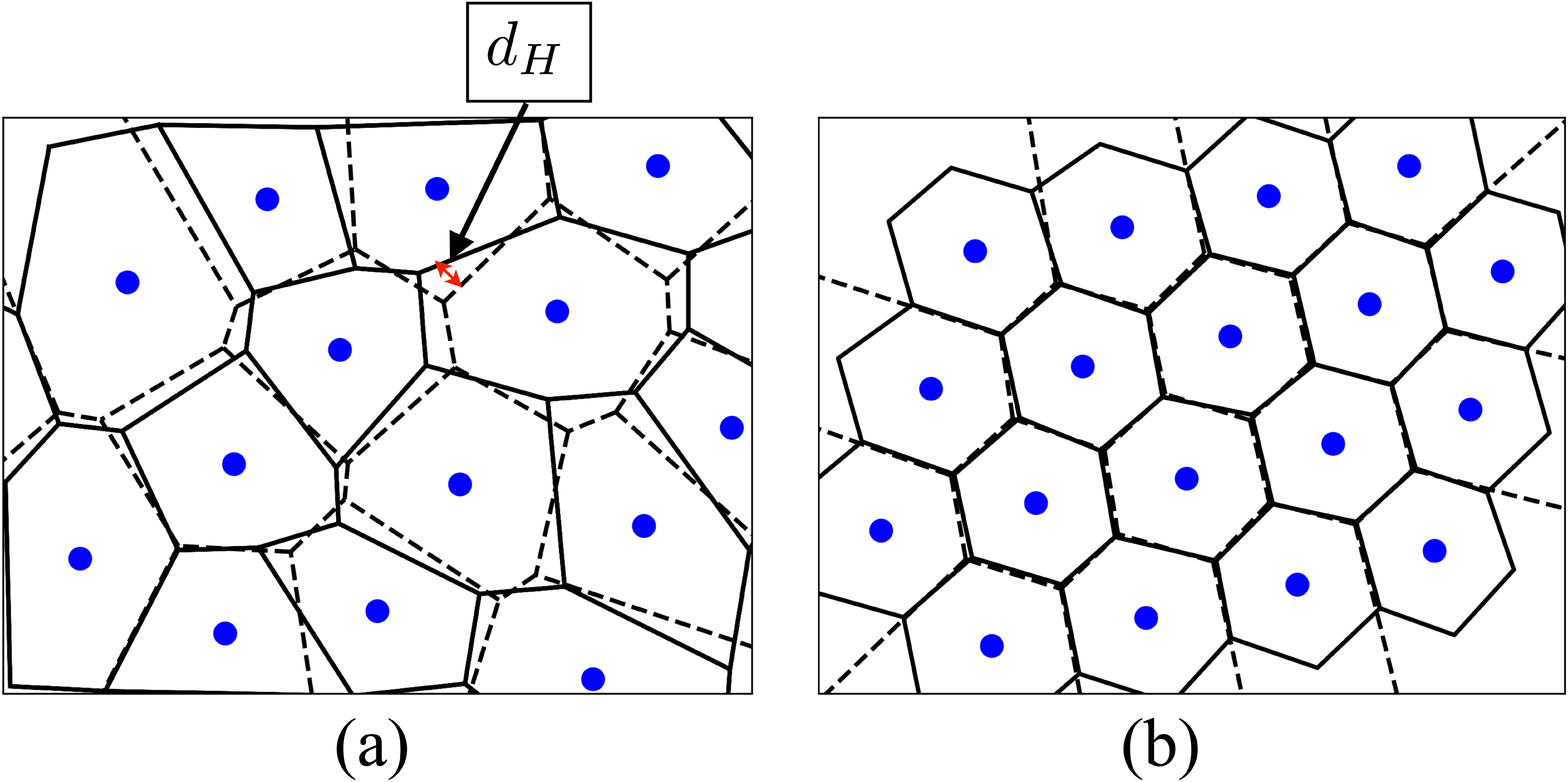}
    \caption{Measuring the Hausdorff Metric, $d_H$. (a) Real mud crack (solid black lines) is compared with the Voronoi diagram (broken black lines) generated using the centroids of the polygons. The red arrow shows where they deviate most. This deviation gives the Hausdorff Metric. (b) The Voronoi diagram matches perfectly with the honeycomb structure. The Hausdorff Metric is zero for them.}
    \label{hausdorff}
\end{figure}

To illustrate the calculation of the Hausdorff Metric, we refer to Fig. \ref{hausdorff}(a) that shows the real mud crack of Case II (solid black lines) and the Voronoi diagram (broken black lines) generated by using the polygon centroids as the seeds. The red arrow shows where the real crack pattern deviates most from its corresponding Voronoi diagram. This distance gives the Hausdorff Metric between the original crack network and its corresponding Voronoi diagram, which becomes zero if the two sets are exactly identical. For the honeycomb, Fig. (\ref{hausdorff}b), the real crack mosaic and its corresponding Voronoi diagram overlap each other perfectly and  the Hausdorff Metric is zero. Examination of Table \ref{tab_quantification} displays that the Hausdorff Metric is zero only for the case of the honeycomb structure. This is not surprising as the honeycomb of Fig. \ref{triangulation}(d) being a regular hexagon, the seed of every hexagon coincides with its centroid.  The Hausdorff Metric has a non-zero value for the mud cracks though the value decreases as the mosaic becomes more Centroidal Voronoi-like. 

In summary, the $p$ measure can certainly provide the information that a crack network is not a Centroidal Voronoi if its $p \ne 0$. The Hausdorff Metric between a polygonal crack network and its corresponding Voronoi diagram (with the centroids of the original crack network as seed) is zero, if and only if the crack network is a Centroidal Voronoi diagram. Lastly even though natural crack mosaics may not all be Centroidal Voronoi, the natural tendency of all systems appears to veer towards a Voronoi diagram to optimize system energy.


Table (\ref{tab_quantification}) indicates the $(n,v,D,\lambda)$ tuple values for each of the static crack mosaics, and their positions in the topology-geometry domain are also indicated in Fig.(\ref{pattern-gil-uni-3d}). The honeycomb mosaic is almost a Centroidal Voronoi but both the mud crack systems are Voronoi-like.

\section{Conclusions}
This work  examines a dynamical crack system developing in a colloidal system,  specifically, the cyclic wetting and drying of mud cracks that lead to an evolving geometry of the mosaic. Colloidal systems in the slurry state are characteristically soft and have long response times. These inherent properties have enabled observation of evolving crack mosaics. The authors have simulated the experimental observations of cyclic wetting-drying on mud systems using a spring model. The evolving crack mosaics have been observed to change shape from `T'-junctions to `Y'-junctions over time, matching the  observations of geologists and physicists reported earlier. The evolution of the topology and geometry of crack mosaic has been mapped as a trajectory onto a $(n,v,D,\lambda)$ domain.

The authors use a phenomenological argument for energy changes that occur during periodic wetting and drying of mud systems, as a function of crack geometry. Energy of the system shows a power-law dependence on $\lambda$ with the exponent value $1.13$. We demonstrated that it is the natural tendency of the system to move towards Voronoi-like tessellations as reported by experimentalists, in order to minimize total system energy. The same conclusions were obtained from examination of the mathematical description of geometrical energy. 

The basic feature that all the space in a Voronoi diagram have the closest proximity to its seed, obviously has a role in this preference of natural systems. The authors develop a quantification of `Voronoi-ness' through a $p$ measure and the Hausdorff Metric, to study the evolution of the dynamical crack mosaic as a function of drying-wetting cycles towards a Voronoi diagram. Our studies indicate that under repeated wetting and drying cycles, mud cracks mature towards Centroidal Voronoi tessellation. Further, our simulations indicate that a system with lower elastic modulus when subjected to longer drying spells between consecutive wetting, is conducive to reach a mature Voronoi mosaic quickly.

Finally  to explore if static crack mosaics also prefer Voronoi patterns, some examples of static crack mosaics from the real world have also been examined and their Voronoi-ness is quantified in terms of the $p$ measure and the Hausdorff Metric  between the original polygonalized crack network and the corresponding Voronoi mosaic. In these systems, the polygonal centroid was assumed to act as the Voronoi seed. The Hausdorff metric was calculated to determine the extent of correlation between the control centre (the polygon centroid) and dynamics.

This study of wetting-drying cycles in mud can help to understand the ‘why and how’ of complex patterns in similar network systems. To the best of our knowledge, this kind of theoretical and simulation study is one of the first efforts in this direction. The quantification of `Voronoi-ness' will help in modeling of tessellations observed both in nature and man-made systems. Understanding the underlying fundamental principles, that of system energy minimization, that drive such tessellations will be aided by our study. This is further facilitated as not only is the geometric theory of Centroidal Voronoi tessellations well developed, but physical tessellations that are close to such Voronoi tessellations, approximately share several of their geometric properties.

\section*{Acknowledgement}
The authors are grateful to Anamika Roy for their helpful discussions and insightful comments. RAI Haque acknowledges support from research fellowships funded by UGC (UGC-ref No. 1435/(CSIR-UGC NET JUNE 2017).

\section*{Appendix: Construction of Voronoi diagram from a seed scatter}\label{voronoi}
\begin{figure}[h!] 
    \centering
    \includegraphics[width=\linewidth]{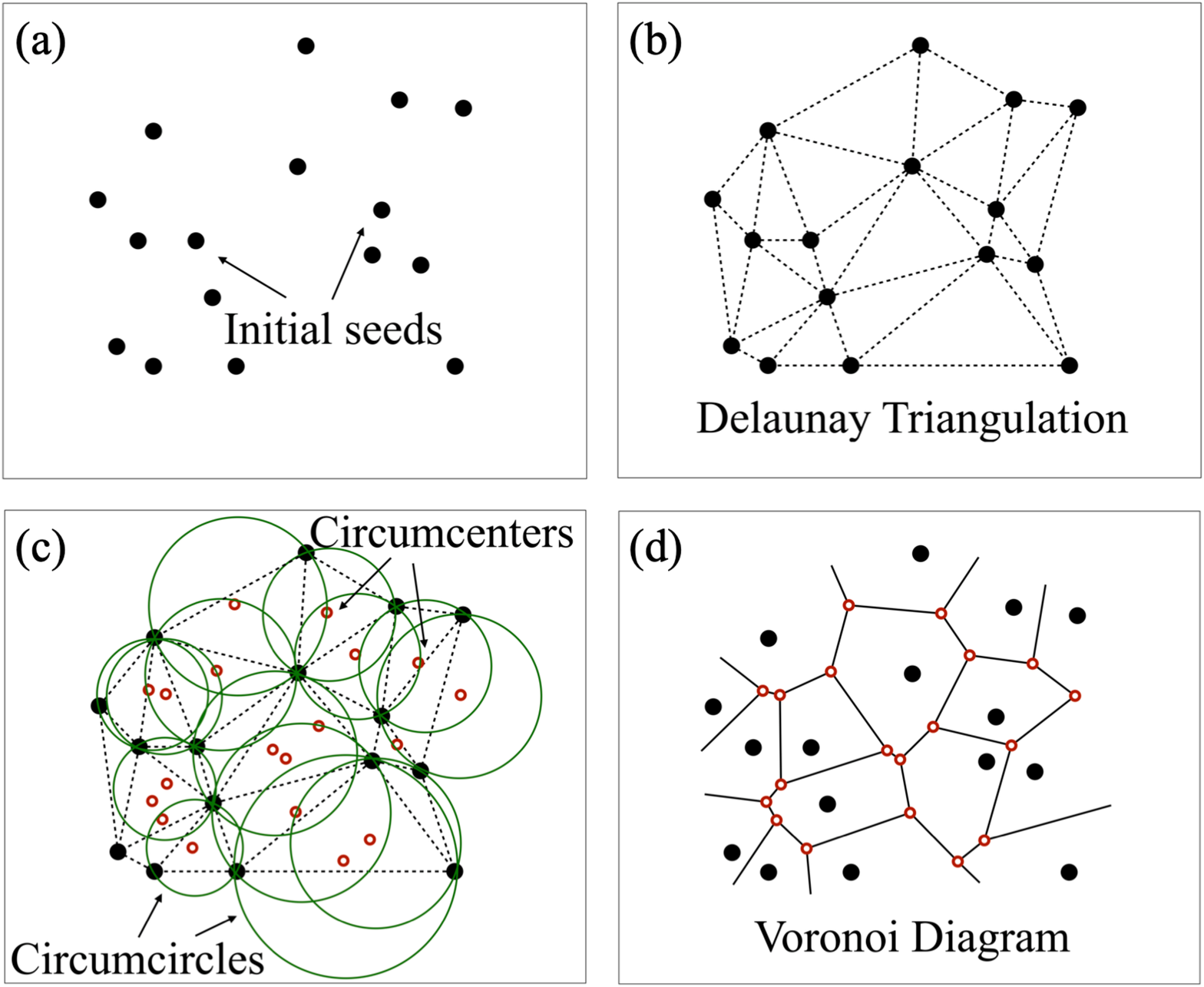} %
    \caption{A Voronoi diagram is the dual of a Delaunay triangulation: (a) Seeds used to generate the Voronoi Diagram, (b) Delaunay triangulation of the seeds. (c) Not a single seed is in the interior of any of the circumcircles (as expected). Red points are the circumcenters of the triangles. (d) Joining the circumcenters of the triangles that share a common side produces the Voronoi diagram.}
    \label{voronoi_dual}
\end{figure}
 Fig. \ref{voronoi_dual} is a schematic flowchart for the generation of Voronoi diagram from a scatter of points (seeds), Fig. \ref{voronoi_dual}(a). The points are connected by lines to their  neighbours to form a Delaunay triangulation, Fig. \ref{voronoi_dual}b; followed by construction of circumcircles of every triangle, Fig. \ref{voronoi_dual}(c); finally joining the circumcenters (marked by red in the online version) of the triangles that share a common side completes the construction of the Voronoi diagram, Fig. \ref{voronoi_dual}(d). If the points coincide with the centroids of the Voronoi polygons, the Voronoi diagram is called a \textit{Centroidal Voronoi} diagram.
 %
\bibliographystyle{elsarticle-num}
\bibliography{ref_wd}

\end{document}